\documentclass[doublecol]{epl2} 

\usepackage{amssymb}
\usepackage{slashed}
\usepackage{hyperref}
\usepackage{lineno,geometry,amsmath,appendix}
\modulolinenumbers[5]
\usepackage{bbold}
\def\s{\mbox{\boldmath$\displaystyle\mathbb{\sigma}$}}

\def\p{\mbox{\boldmath$\displaystyle\boldsymbol{p}$}}
\def\0{\mbox{\boldmath$\displaystyle\boldsymbol{0}$}}

\def\x{\mbox{\boldmath$\displaystyle\boldsymbol{x}$}}
\def\y{\mbox{\boldmath$\displaystyle\boldsymbol{y}$}}

\title{Equivalence of regular spinor fields}
\author{Cheng-Yang Lee \thanks{Email: \email{cylee@scu.edu.cn}}}

\institute{                    
 Center for Theoretical Physics, College of Physics,\\
Sichuan University, Chengdu, 610064, China.

}

\abstract{
In the Lounesto classification, there are three types of regular spinors. They are classified by the condition that at least one of the scalar or pseudo scalar norms are non-vanishing. The Dirac spinors are regular spinors because their scalar and pseudo scalar norms are non-zero and zero respectively. We construct local and Lorentz-covariant fermionic fields from all three classes of regular spinors. By computing the invariants and bilinear covariants of the regular spinor fields, we show that they are physically equivalent to the Dirac fields in the sense that whatever interactions one writes down using the regular spinor fields, they can always be expressed in terms of the Dirac fields.
}

\begin{document}

\maketitle

\section{Introduction}

Ever since Dirac wrote down the equation of motion describing the electrons~\cite{Dirac:1928hu}, spinors have played indispensable roles in quantum field theory and particle physics. Given the importance of spinors, it is imperative to understand their mathematics and 
physics in totality.

Starting from the works of Ahluwalia and Grumiller~\cite{Ahluwalia:2004ab,Ahluwalia:2004sz} followed by da Rocha and Rodrigues~\cite{daRocha:2005ti}, the Lounesto classification of spinors~\cite{Lounesto:2001zz} have received much
attention~\cite{HoffdaSilva:2012uke,daRocha:2013qhu,Cavalcanti:2014wia,Bonora:2014dfa,Fabbri:2017lvu,HoffdaSilva:2017waf,Bonora:2017oyb,Lee:2018ull,Arcodia:2019flm,Fabbri:2020elt,Rogerio:2020ewe,Rogerio:2023cwv,Papadopoulos:2024ouk}. According to the classification, the Dirac and Weyl spinors which describe massive 
and massless fermions in the Standard Model, do not encompass all the possibilities. In the $(\frac{1}{2},0)\oplus(0,\frac{1}{2})$ representation of the Lorentz group, the classification utilizes the following bilinear covariants of the spinor $\lambda$ and its dual $\overline{\lambda}$
\begin{equation}
\begin{array}{ll}
\Omega_{1}=\overline{\lambda}\lambda, &\Omega_{2}=\overline{\lambda}\gamma^{5}\lambda,\\
K^{\mu}=\overline{\lambda}\gamma^{5}\gamma^{\mu}\lambda,&
 S^{\mu\nu}=\overline{\lambda}\gamma^{\mu}\gamma^{\nu}\lambda,
\end{array}
\end{equation}
where $\overline{\lambda}=\lambda^{\dag}\gamma^{0}$ and
\begin{align}
\gamma^{0}&=
\left(\begin{array}{cc}
\mathbb{O} & \mathbb{I} \\
\mathbb{I} & \mathbb{O}
\end{array}\right),\,
\gamma^{i}=
\left(\begin{array}{cc}
\mathbb{O} & -\sigma^{i} \\
\sigma^{i} & \mathbb{O}
\end{array} \right),\\
\gamma^{5}&=
\left(\begin{array}{cc}
\mathbb{I} & \mathbb{O} \\
\mathbb{O} &-\mathbb{I} \end{array}
\right).\label{eq:gamma}
\end{align}
The Lounesto classification is given by
\begin{enumerate}
\item $\Omega_{1}\neq0$, $\Omega_{2}\neq 0$
\item $\Omega_{1}\neq0$, $\Omega_{2}=0$
\item $\Omega_{1}=0, \Omega_{2}\neq 0$
\item $\Omega_{1}=0,$ $\Omega_{2}=0$, $K^{\mu}\neq0$, $S^{\mu\nu}\neq0$
\item $\Omega_{1}=0,$ $\Omega_{2}=0$, $K^{\mu}= 0$, $S^{\mu\nu}\neq0 $
\item $\Omega_{1}=0,$ $\Omega_{2}=0$, $K^{\mu}\neq 0$, $S^{\mu\nu}=0$
\end{enumerate}
In this paper, we have adopted the notation that the top and bottom components of the four-component spinors to be the right- and left-handed Weyl spinors respectively. Spinors belonging to the first three classes are called \textit{regular spinors} while those belonging to the remaining classes are called \textit{singular spinors}. The Dirac and Weyl spinors belong to the 2nd and 6th class respectively. By Dirac and Weyl spinors, we mean spinors that satisfy $(\gamma^{\mu}p_{\mu}\pm m\mathbb{I})\lambda=0$ and $\gamma^{\mu}p_{\mu}\lambda'=0$ respectively.\footnote{In the physics literature, Weyl spinors which describe massless fermions are the left and right-handed two-component spinors. For the purpose of the Lounesto classification, these two-component Weyl spinors are expressed in terms of four-component spinors satisfying the massless Dirac equation in the momentum space.} By solving these equations, one finds $\lambda$ and $\lambda'$ to reside in the 2nd and 6th class respectively.

A natural question that arises from the classification is - \textit{What are the physics of the remaining classes of spinors?} In this work, we focus on the regular spinors belonging to the 1st-3rd class (For the physics of singular spinors belonging to the 4th and 5th class, please see~\cite{Ahluwalia:2022ttu,Ahluwalia:2022yvk,Rogerio:2023cwv,Ahluwalia:2023slc}). 

%In the literature, there are claims that quantum fields constructed from different classes of regular spinors are physically distinct. 

%Our purpose is to refute these claims. 

Following the formalism developed by Wigner and Weinberg~\cite{Wigner:1939cj,Weinberg:1995mt}, we construct the most general massive spin-half quantum fields in the $\left(\frac{1}{2},0\right)\oplus\left(0,\frac{1}{2}\right)$ representation. Contrary to Weinberg, we find that the spinors cannot be completely fixed by locality and parity symmetry. That is, there remains freedom in the choice of phases for the spinors. By choosing appropriate values of the phases, we obtain local and Lorentz-covariant fermionic fields for all three classes of regular spinors.

The main result of this paper is that all fermionic fields constructed from regular spinors are physically equivalent to the Dirac fields. This is accomplished by showing that they are all related to the Dirac fields via unitary transformations. Similarly, all bilinear covariants constructed from regular spinor fields can be expressed in terms of the bilinear covariants of the Dirac field. Therefore, whatever interaction one writes down using the regular spinor fields, we can always write down a physically equivalent counterpart in terms of the Dirac fields.

%The paper is organized as follows. In sec.~\ref{sec:reg}, we construct the spin-half quantum fields using regular spinors and establish their physical equivalence. We conclude in sec.~\ref{conc} by discussing the consequences when the demand of Hermiticity is relaxed.

%------------------------------------------------------
\section{Regular spinor fields}
%------------------------------------------------------

Let $\psi$ and $\overline{\psi}$ be the massive spin-half field and its dual in the $\left(\frac{1}{2},0\right)\oplus\left(0,\frac{1}{2}\right)$ representation. With the appropriate normalizations, their expansions are given by
\begin{align}
\psi(x)=(2\pi)^{-3/2}\int&\frac{d^{3}p}{\sqrt{2E}}\sum_{\sigma}\big[e^{-ip\cdot x}u_{\sigma}(\p)a_{\sigma}(\p)\nonumber\\
&+e^{ip\cdot x}v_{\sigma}(\p)b^{\dag}_{\sigma}(\p)\big].
\end{align}
\begin{align}
\overline{\psi}(x)=(2\pi)^{-3/2}\int&\frac{d^{3}p}{\sqrt{2E}}\sum_{\sigma}\big[e^{ip\cdot x}\overline{u}_{\sigma}(\p)a^{\dag}_{\sigma}(\p)\nonumber\\
&+e^{-ip\cdot x}\overline{v}_{\sigma}(\p)b_{\sigma}(\p)\big].
\end{align}
We restrict the quantum fields to furnish the irreducible representations of the Poincar\'{e} group without introducing additional degeneracies so $\sigma=\pm\frac{1}{2}$. In the rest frame, rotational symmetry constrain the spinors to take the form~\cite{Weinberg:1995mt}
\begin{align}
u_{\frac{1}{2}}&=\sqrt{m}\left(\begin{matrix}
c_{+} \\
0\\
c_{-}\\
0
\end{matrix}\right),\,
u_{-\frac{1}{2}}=\sqrt{m}\left(\begin{matrix}
0 \\
c_{+} \\
0\\
c_{-}
\end{matrix}\right),\label{eq:uc}\\
v_{\frac{1}{2}}&=\sqrt{m}\left(\begin{matrix}
0\\
d_{+} \\
0\\
d_{-}\\
\end{matrix}\right),\,
v_{-\frac{1}{2}}=-\sqrt{m}\left(\begin{matrix}
d_{+} \\
0\\
d_{-} \\
0
\end{matrix}\right),\label{eq:vd}
\end{align}
where $c_{\pm}$ and $d_{\pm}$ are non-vanishing constants to be determined. Spinors at arbitrary momentum are obtained through the boost
\begin{align}
u_{\sigma}(\p)&=D(\p)u_{\sigma},\\
v_{\sigma}(\p)&=D(\p)v_{\sigma},
\end{align}
where
\begin{equation}
	D(\p)=\sqrt{\frac{E+m}{2m}}\left[\begin{array}{cc}
		\mathbb{I}+\frac{\boldsymbol{\sigma\cdot p}}{E+m}&\mathbb{O}\\
		\mathbb{O}&\mathbb{I}-\frac{\boldsymbol{\sigma\cdot p}}{E+m}
	\end{array}\right]
\end{equation}
and $E=\sqrt{|\p|^{2}+m^{2}}$. The invariants of the spinors evaluate to
\begin{align}
\overline{u}_{\sigma}(\p)u_{\sigma'}(\p)&=m(c^{*}_{-}c_{+}+c_{-}c^{*}_{+})\delta_{\sigma\sigma'}, \\
\overline{u}_{\sigma}(\p)\gamma^{5}u_{\sigma'}(\p)&=m(c^{*}_{-}c_{+}-c_{-}c^{*}_{+})\delta_{\sigma\sigma'},
\end{align}
and
\begin{align}
\overline{v}_{\sigma}(\p)v_{\sigma'}(\p)&=m(d^{*}_{-}d_{+}+d_{-}d^{*}_{+})\delta_{\sigma\sigma'},\\
\overline{v}_{\sigma}(\p)\gamma^{5}v_{\sigma'}(\p)&=m(d_{-}d^{*}_{+}-d^{*}_{-}d_{+})\delta_{\sigma\sigma'}.
\end{align}
Since $c_{\pm}$ and $d_{\pm}$ are non-vanishing, by appropriate choice of their values, $u_{\sigma}$ and $v_{\sigma}$ encompass all the regular spinors. The Lounesto classification for the $u$ spinors are given by
\begin{enumerate}
\item $\Omega_{1}\neq 0$, $\Omega_{2}\neq0$, $c^{*}_{-}c_{+}\neq\pm c_{-}c^{*}_{+}$,
\item $\Omega_{1}\neq 0$, $\Omega_{2}=0$, $c^{*}_{-}c_{+}=+c_{-}c^{*}_{+}$, 
\item $\Omega_{1}=0$, $\Omega_{2}\neq0$, $c^{*}_{-}c_{+}=-c_{-}c^{*}_{+}$.
\end{enumerate}
Similarly for the $v$ spinors, we have
\begin{enumerate}
\item $\Omega_{1}\neq 0$, $\Omega_{2}\neq0$, $d^{*}_{-}d_{+}\neq\pm d_{-}d^{*}_{+}$,
\item $\Omega_{1}\neq 0$, $\Omega_{2}=0$, $d^{*}_{-}d_{+}=+d_{-}d^{*}_{+}$, 
\item $\Omega_{1}=0$, $\Omega_{2}\neq0$, $d^{*}_{-}d_{+}=-d_{-}d^{*}_{+}$.
\end{enumerate}

The classification of $u$ and $v$ raises an interesting question. For the spinor fields $\psi$ and $\overline{\psi}$, can $u$ and $v$ belong to different classes? This turns out to be impossible. The demand of locality impose further constraints on the coefficients, forcing $u$ and $v$ to to be in the same class. To see this, we compute the equal-time anti-commutator/commutator for $\psi$ and $\overline{\psi}$ 
\begin{align}
\left[\psi(t,\x),\overline{\psi}(t,\y)\right]_{\pm}=&(2\pi)^{-3}\int\frac{d^{3}p}{2E}e^{i\boldsymbol{p\cdot(x-y)}}\nonumber\\
&\times\left[N(\p)\pm M(-\p)\right],
\end{align}
where
\begin{align}
N(\p)&=\sum_{\sigma}u_{\sigma}(\p)\overline{u}_{\sigma}(\p) \nonumber\\
&=m\left(\begin{matrix}
c^{*}_{-}c_{+}\mathbb{I} & |c_{+}|^{2}e^{-\boldsymbol{\sigma\cdot\varphi}} \\
|c_{-}|^{2}e^{\boldsymbol{\sigma\cdot\varphi}} & c_{-}c^{*}_{+}\mathbb{I}
\end{matrix}\right),
\end{align}
and
\begin{align}
M(\p)&=\sum_{\sigma}v_{\sigma}(\p)\overline{v}_{\sigma}(\p) \nonumber\\
&=m\left(\begin{matrix}
d^{*}_{-}d_{+}\mathbb{I} & |d_{+}|^{2}e^{-\boldsymbol{\sigma\cdot\varphi}} \\
|d_{-}|^{2}e^{\boldsymbol{\sigma\cdot\varphi}} & d_{-}d^{*}_{+}\mathbb{I}
\end{matrix}\right).
\end{align}
Locality requires the commutator or anti-commutator to be proportional to $\delta^{3}(\x-\y)$. Using the identity
\begin{equation}
e^{\boldsymbol{\sigma\cdot\varphi}}=m^{-1}\left(\sqrt{|\p|^{2}+m^{2}}\mathbb{I}+\s\cdot\p\right),
\end{equation}
we find that the fields can only anti-commute at equal-time and this requires
\begin{equation}
N(\p)+M(-\p)=2E
\left(\begin{matrix}
\mathbb{O} & |c_{+}|^{2}\mathbb{I} \\
|c_{-}|^{2}\mathbb{I} & \mathbb{O}
\end{matrix}\right).
\end{equation}
Therefore, the coefficients must satisfy the following constraints
\begin{equation}
|c_{\pm}|^{2}=|d_{\pm}|^{2},\, c^{*}_{-}c_{+}=-d^{*}_{-}d_{+}.\label{eq:coeff_constr}
\end{equation}
From~(\ref{eq:coeff_constr}), we see that locality demands $u$ and $v$ to be in the same regular class.

The equations of motion for the spinors and field can now be derived. Using the orthonormality relations, we find
\begin{align}
N(\p)u_{\sigma}(\p)&=m(c^{*}_{-}c_{+}+c_{-}c^{*}_{+})u_{\sigma}(\p), \label{eq:N}\\
M(\p)v_{\sigma}(\p)&=-m(c^{*}_{-}c_{+}+c_{-}c^{*}_{+})v_{\sigma}(\p).\label{eq:M}
\end{align}
Substituting the spin-sums into~(\ref{eq:N}-\ref{eq:M}) and simplify, we obtain the equations of motion for the spinors in the momentum space
\begin{align}
(\Gamma^{\mu}p_{\mu}-M)u_{\sigma}(\p)&=0, \\
(\Gamma^{\mu}p_{\mu}+M)v_{\sigma}(\p)&=0,
\end{align}
where
\begin{align}
\Gamma^{\mu}p_{\mu}&=m\left(\begin{matrix}
\mathbb{O} & |c_{+}|^{2}e^{-\boldsymbol{\sigma\cdot\varphi}} \\
|c_{-}|^{2}e^{\boldsymbol{\sigma\cdot\varphi}} & \mathbb{O}
\end{matrix}\right), \\
M&=m\left(\begin{matrix}
c_{-}c^{*}_{+}\mathbb{I} & \mathbb{O} \\
\mathbb{O} & c^{*}_{-}c_{+}\mathbb{I}
\end{matrix}\right),
\end{align}
so the Lagrangian density for the fields are
\begin{equation}
\mathcal{L}=\overline{\psi}\left(i\Gamma^{\mu}\partial_{\mu}-M\right)\psi.
\end{equation}
The conjugate momentum for $\psi$ is $\pi=i\overline{\psi}\Gamma^{0}$ so their equal-time anti-commutator evaluate to
\begin{equation}
\left[\psi(t,\x),\pi(t,\y)\right]_{+}=i|c_{-}c_{+}|^{2}\delta^{3}(\x-\y)I.
\end{equation}
Therefore, we must have $|c_{-}c_{+}|=|d_{-}d_{+}|=1$.

To simplify the analysis, we use the fact that $\psi$ and $\overline{\psi}$ are only defined up to a constant global phase so without the loss of generality, we can choose $c_{+}=1$ and take
\begin{equation}
d_{+}=e^{i\alpha_{1}},\,d_{-}=-e^{i\alpha_{2}}, c_{-}=e^{i(\alpha_{2}-\alpha_{1})}.\label{cd}
\end{equation}
Substituting~(\ref{eq:uc}) into~(\ref{eq:vd}), we obtain
\begin{align}
u_{\frac{1}{2}}&=\sqrt{m}\left[\begin{matrix}
1 \\
0 \\
e^{i(\alpha_{2}-\alpha_{1})}\\
0
\end{matrix}\right],
u_{-\frac{1}{2}}=\sqrt{m}\left[\begin{matrix}
0 \\
1 \\
0 \\
e^{i(\alpha_{2}-\alpha_{1})}
\end{matrix}\right],\\
v_{\frac{1}{2}}&=\sqrt{m}\left[\begin{matrix}
0 \\
e^{i\alpha_{1}} \\
0 \\
-e^{i\alpha_{2}}
\end{matrix}\right],
v_{-\frac{1}{2}}=-\sqrt{m}\left[\begin{matrix}
e^{i\alpha_{1}} \\
0 \\
-e^{i\alpha_{2}}\\
0
\end{matrix}\right].
\end{align}
Using these choice of phases, the Lagrangian density and the propagator take the form
\begin{equation}
\mathcal{L}_{\psi}=\overline{\psi}\left[i\gamma^{\mu}\partial_{\mu}-e^{i\gamma^{5}(\alpha_{2}-\alpha_{1})}m\right]\psi,
\end{equation}
and
\begin{align}
S_{\psi}(x-y)=&i\int\frac{d^{4}q}{(2\pi)^{4}}e^{-iq\cdot(x-y)}\nonumber\\
&\times\left[\frac{\gamma^{\mu}q_{\mu}+e^{i\gamma^{5}(\alpha_{2}-\alpha_{1})}m}{q^{2}-m^{2}+i\epsilon}\right].
\end{align}

Contrary to Weinberg's construction of the Dirac fields which made use of parity, we have only made use of continuous symmetries. The difference between the two constructs can be attributed to the fact that Weinberg took the parity operator in the $\left(\frac{1}{2},0\right)\oplus\left(0,\frac{1}{2}\right)$ representation to be $\gamma^{0}$. However, this choice is not unique. From the extended Lorentz algebra, the parity operator $P$ is a linear operator that commutes and anti-commutes with the rotation and boost generators. Here, the rotation and boost generators are
\begin{equation}
\boldsymbol{\mathcal{J}}=\frac{1}{2}\left(\begin{matrix}
\s & \mathbb{O} \\
\mathbb{O} & \s
\end{matrix}\right)\label{eq:j}
\end{equation}
and
\begin{equation}
\boldsymbol{\mathcal{K}}=\frac{1}{2}\left(\begin{matrix}
i\s & \mathbb{O} \\
\mathbb{O} & -i\s
\end{matrix}\right)\label{eq:k}
\end{equation}
respectively. Using~(\ref{eq:j}-\ref{eq:k}), we find
\begin{equation}
P=\left(\begin{matrix}
\mathbb{O} & \beta_{2}\mathbb{I} \\
\beta_{1}\mathbb{I} & \mathbb{O}
\end{matrix}\right)
\end{equation}
where $\beta_{1,2}$ are arbitrary constants. For any regular spinor fields, parity is conserved. Taking
\begin{align}
U(P)a_{\sigma}(\p)U^{-1}(P)&=\eta^{*}a_{\sigma}(-\p),\\
U(P)b^{\dag}_{\sigma}(\p)U^{-1}(P)&=\overline{\eta}b^{\dag}_{\sigma}(-\p),
\end{align}
where $\eta$ and $\overline{\eta}$ are the intrinsic parity for particle and anti-particle, we find
\begin{equation}
U(P)\psi(t,\x)U^{-1}(P)=\eta^{*}P\psi(t,-\x)
\end{equation}
provided that we take the intrinsic parity to be odd $\eta^{*}=-\overline{\eta}$ and choose the phases to be
\begin{equation}
\beta_{1}=\beta^{-1}_{2}=e^{i(\alpha_{2}-\alpha_{1})}.
\end{equation}

Having obtained the Lagrangian density and propagator for the regular spinor fields,
we can construct interactions using $\psi$ and $\overline{\psi}$. By choosing
different values of $\alpha_{1}$ and $\alpha_{2}$, the scattering amplitudes associated with different classes of regular spinors will in general, be different. This seems to suggest that different regular
spinors entail different physics but this is not true. In what is to
follow, we will show that whatever interactions one constructs in terms of $\overline{\psi}$ and $\psi$, they can always be expressed in terms of the Dirac fields. Consequently, the physics of regular spinors
are encoded within the Dirac fields. To see this, let us take the Dirac fields to be
\begin{align}
\chi=(2\pi)^{-3/2}\int&\frac{d^{3}p}{\sqrt{2E}}\sum_{\sigma}\Big[e^{-ip\cdot x}\xi_{\sigma}(\p)a_{\sigma}(\p)\nonumber\\
&+e^{ip\cdot x}\zeta_{\sigma}(\p)b^{\dag}_{\sigma}(\p)\Big]
\end{align}
and $\overline{\chi}=\chi^{\dag}\gamma^{0}$ where
\begin{align}
\xi_{\frac{1}{2}}&=\sqrt{m}\left(\begin{matrix}
1 \\
0 \\
1 \\
0
\end{matrix}\right),
&\xi_{-\frac{1}{2}}&=\sqrt{m}\left(\begin{matrix}
0 \\
1 \\
0 \\
1
\end{matrix}\right),\\
\zeta_{\frac{1}{2}}&=\sqrt{m}
\left(\begin{matrix}
0 \\
e^{i\alpha_{1}} \\
0 \\
-e^{i\alpha_{1}}
\end{matrix}\right),
&\zeta_{-\frac{1}{2}}&=\sqrt{m}\left(\begin{matrix}
-e^{i\alpha_{1}} \\
0 \\
e^{i\alpha_{1}} \\
0
\end{matrix}\right).
\end{align}
The global phase $e^{i\alpha_{1}}$ is allowed by rotation symmetry. From $\xi$ and $\zeta$, we see that
they belong to the 2nd class. The Lagrangian density and propagator for the Dirac fields are
\begin{equation}
\mathcal{L}_{\chi}=\overline{\chi}\left(i\gamma^{\mu}\partial_{\mu}-m\mathbb{I}\right)\chi
\end{equation}
and
\begin{align}
S_{\chi}=&i\int\frac{d^{4}q}{(2\pi)^{4}}e^{-iq\cdot(x-y)}
\left[\frac{\gamma^{\mu}q_{\mu}+m\mathbb{I}}{q^{2}-m^{2}+i\epsilon}\right].
\end{align}
The relations between $\psi,\overline{\psi}$ and $\chi,\overline{\chi}$ are
\begin{align}
\psi(x)&=A\chi(x),\label{eq:psi_chi1}\\
\overline{\psi}(x)&=\overline{\chi}(x)B,\label{eq:psi_chi2}
\end{align}
where
\begin{align}
A&=
\left[\begin{matrix}
\mathbb{I} & \mathbb{O} \\
\mathbb{O} & e^{i(\alpha_{2}-\alpha_{1})}\mathbb{I}
\end{matrix}\right], \\
B&=
\left[\begin{matrix}
e^{-i(\alpha_{2}-\alpha_{1})}\mathbb{I} & \mathbb{O} \\
\mathbb{O} & \mathbb{I} 
\end{matrix}\right],
\end{align}
from which we find $\mathcal{L}_{\psi}=\mathcal{L}_{\chi}$
and
\begin{equation}
S_{\psi}(x-y)=AS_{\chi}(x-y)B.
\end{equation}
The invariants and covariants of $\psi$ and $\chi$ are related by
\begin{align}
\overline{\psi}\psi&=\overline{\chi}e^{-i\gamma^{5}(\alpha_{2}-\alpha_{1})}\chi,\label{eq:scalar}\\
\overline{\psi}\gamma^{5}\psi&=\overline{\chi}\gamma^{5}e^{-i\gamma^{5}(\alpha_{2}-\alpha_{1})}\chi,\label{eq:pseudo_scalar}\\
\overline{\psi}\gamma^{\mu}\psi&=\overline{\chi}\gamma^{\mu}\chi, \\
\overline{\psi}\gamma^{\mu}\gamma^{5}\psi&=\overline{\chi}\gamma^{\mu}\gamma^{5}\chi, \\
\overline{\psi}\gamma^{\mu}\gamma^{\nu}\psi&=\overline{\chi}e^{-i\gamma^{5}(\alpha_{2}-\alpha_{1})}\gamma^{\mu}\gamma^{\nu}\chi.\label{eq:tensor}
\end{align}
By construction, we know that the covariants and invariants of $\psi$ and $\overline{\psi}$ transforms as a scalar field under continuous Lorentz transformations. However, because the parity operator associated with $\psi$ and $\overline{\psi}$ is not $\gamma^{0}$ when $\alpha_{1}\neq\alpha_{2}$, so how these quantities transform under parity are different to the Dirac fields $\chi$ and $\overline{\chi}$. For example, from~(\ref{eq:scalar}-\ref{eq:pseudo_scalar}), we see that $\overline{\psi}\psi$ and $\overline{\psi}\gamma^{5}\psi$ are in general, neither scalar and pseudo-scalar under parity. To determine how these quantities transform, it is convenient to express the bilinears of the Dirac fields in terms of $\overline{\psi}$ and $\psi$. Direct calculations yield
\begin{align}
\overline{\chi}\chi&=\overline{\psi}e^{i\gamma^{5}(\alpha_{2}-\alpha_{1})}\psi,\\
\overline{\chi}\gamma^{5}\chi&=\overline{\psi}\gamma^{5}e^{i\gamma^{5}(\alpha_{2}-\alpha_{1})}\psi,\\
\overline{\chi}\gamma^{\mu}\gamma^{\nu}\chi&=\overline{\psi}e^{i\gamma^{5}(\alpha_{2}-\alpha_{1})}\gamma^{\mu}\gamma^{\nu}\psi.
\end{align}
In this way, we obtain the correct scalar, pseudo-scalar and tensor for $\overline{\psi}$ and $\psi$.

The demand of Lorentz symmetry means that interactions ought to be constructed from
the invariants and covariants of $\overline{\psi}$ and $\psi$. The most common ones are given in (\ref{eq:scalar}-\ref{eq:tensor})
and they can always be written in terms of the Dirac field and its adjoint.
One can construct other invariants and covariants but because of (\ref{eq:psi_chi1}-\ref{eq:psi_chi2}), we can always write the interactions in terms of $\overline{\chi}$ and $\chi$.

\section{Conclusions}\label{conc}

The Lounesto classification has inspired physicists to study theoretical constructs beyond the Dirac and Weyl fields with potentially important consequences for physics beyond the Standard Model. 

For singular spinors residing in the 4th-5th class, their quantum fields are physically distinct from the Dirac and Weyl fields which reside in the 2nd and 6th class. The important question is - what are the physics of the regular spinors in the 1st and 3rd class. 

In the literature, there are claims that different classes of regular spinors entail different physics. Here, we show that these claims are incorrect. For this purpose, we have constructed massive fermionic fields from all three classes of regular spinors. By comparing their Lagrangian densities, propagators and bilinear covariants, we find that these quantities can all be expressed in terms of the Dirac field and its adjoint. Consequently, any interactions constructed using quantum fields constructed from regular spinors in the 1st and 3rd class, they can always be expressed in terms of the Dirac fields. Therefore, as far as quantum field theory is concerned, all regular spinors are physically equivalent.

%With this work, we believe that the physics of regular spinors is settled up to an
%important caveat to be discussed now. In constructing the quantum fields, we have made the assumption that the theory is Hermitian. Specifically, the invariants and covariants
%constructed from $\overline{\psi}$ and $\psi$ are Hermitian. Within the framework of Hermitian Hamiltonian,
%the result presented here is valid. However, if we relax the demand of Hermiticity and
%allow the Hamiltonian to be pseudo Hermitian [14, 15], something remarkable happens.
%The quantum fields, constructed from regular spinors with pseudo Hermitian adjoint are
%bosonic [12, 16]. That is, regular spinors can be used to construct bosonic as well as
%fermionic fields. The analysis of sec. 3 can be readily applied to the pseudo Hermitian
%bosonic fields. The result is easily anticipated - bosonic fields constructed from the three
%classes of regular spinors are physically equivalent.
\acknowledgments
I would like to thank Dharam Vir Ahluwalia, Julio M. Hoff da Silva and R.~J.~Bueno Rogerio for useful discussions. I am grateful to the referee's comments which lead to improvements of the manuscript.

\bibliography{Bibliography}
\bibliographystyle{eplbib.bst}

%\begin{thebibliography}{0}
%
%\bibitem{b.a}
%  \Name{Author F., Author S. \and Author T.}
%  \REVIEW{Some Rev. A}{69}{1969}{9691}.
%
%\bibitem{b.b}
%  \Name{Author F. \and Author S.}
%  \Book{Some Book of Interest}
%  \Editor{A. Editor}
%  \Vol{9}
%  \Publ{Publishing house, City}
%  \Year{1939}
%  \Page{666}.
%
%\bibitem{b.c}
%  \Editor{Editor A.}
%  \Book{Some Book of Interest}
%  \Vol{9}
%  \Publ{Publishing house, City}
%  \Year{1939}
%  \Section{A}.
%
%\end{thebibliography}

\end{document}